\documentclass{PoS}

\title{$D\bar{D}$ and $DD$ pair production at the LHCb in the parton Reggeization approach}

\ShortTitle{$D\bar{D}$ and $DD$ pair production at the LHCb in the parton Reggeization approach}

\author{\speaker{A.V. Karpishkov}%
         \thanks{The work of A.~V.~Karpishkov, M.A.~Nefedov, V.A.~Saleev and A.~V.~Shipilova was
supported in part by the Russian Foundation for Basic Research
through the Grant No. 14-02-00021.}\\
        Samara National Research University, Moscow
Highway, 34, 443086, Samara, Russia\\
        E-mail: \email{karpishkov@rambler.ru}}

\author{M.A. Nefedov\\
         Samara National Research University, Moscow
Highway, 34, 443086, Samara, Russia\\
         E-mail: \email{nefedovma@gmail.com}}

\author{V.A. Saleev\\
         Samara National Research University, Moscow
Highway, 34, 443086, Samara, Russia\\
         E-mail: \email{saleev@samsu.ru}}

\author{A.V. Shipilova\\
         Samara National Research University, Moscow
Highway, 34, 443086, Samara, Russia\\
         E-mail: \email{alexshipilova@samsu.ru}}

\abstract{We study the inclusive $D\overline{D}$ and $DD$ pair
production in proton-proton collisions at the LHC at leading order
of the parton Reggeization approach endowed with universal
scale-depended fragmentation functions for $c$-quark to $D$-meson and
for gluon to $D$-meson transitions. We have described $D\bar D$ and
$DD$ distributions in azimuthal angle, as well as transverse
momentum, rapidity distance, and invariant mass measured in the
region of large rapidity $2 < y < 4$ by the LHCb Collaboration at the LHC without free parameters. We have used Reggeized amplitudes
for the processes $RR\to gg$ and $RR\to c\bar c$ which are obtained
accordingly to Feynman rules of the L.N.~Lipatov effective theory of
Reggeized partons, and Kimber-Martin-Ryskin model for unintegrated
gluon distribution function in a proton with
Martin-Stirling-Thorne-Watt collinear parton distributions as
inputs.}

\FullConference{XXIV International Workshop on Deep-Inelastic Scattering and Related Subjects\\
                  11-15 April, 2016\\
                  DESY Hamburg, Germany}

\begin{document}

\section{Introduction}

The study of the $D$-meson production at high energies provides a crucial test of the next-to-leading order (NLO) calculations in perturbative quantum chromodynamics (QCD)
due to the smallness of strong coupling constant $\alpha_S(\mu)$, as the lowest limit of typical
energy scale of the hard interaction $\mu$ is controlled by the charm quark mass $m\gg\Lambda_{QCD}$, where $\Lambda_{QCD}$
is the asymptotic scale parameter of QCD. At the same time, we achieve a new dynamical regime, ~\textit{Regge limit}, where $\sqrt{S}\gg
\mu\gg\Lambda_{QCD}$ and the large coefficients of new type $\log^n(\sqrt S/\mu)$ should be resummed in all-order terms of perturbative QCD series,
introducing a new small parameter $x=\mu/\sqrt S$. For this purpose we develop the $k_T$-factorization framework~\cite{KTCollins,KTGribov,KTCatani}
endowed with the fully gauge-invariant amplitudes with \textit{Reggeized} gluons in the initial state and call this combination the Parton Reggeization Approach (PRA).

Recently, we demonstrated the
advantages of the PRA in studies of
bottom-flavored jets
\cite{bbTEV,bbLHC}, charmonium and bottomonium production
\cite{KniehlSaleevVasin1,KniehlSaleevVasin2,PRD2003,NSS_charm,NSS_bot}, inclusive
production of single jet \cite{KSSY}, pair of jets \cite{NSSjets},
prompt-photon \cite{tevatronY,heraY}, photon plus jet \cite{KNS14},
Drell-Yan lepton pairs \cite{NNS_DY}.

\section{Basic Formalism}
\label{sec:two}

We study the pair production of $D$-mesons with high transverse momenta
using the factorization theorem of QCD
for the $D\overline{D}$-cross section:
\begin{eqnarray}
\frac{d\sigma(p+p\to D_i(p_D)+\overline{D}_j(p_{\overline{D}})+X)}{dp_{TD} dy_D dp_{T\overline{D}} dy_{\overline{D}}}=
\sum\limits_{i\ j} \int\limits_0^1 \!\! \frac{dz_1}{z_1}D_{i\to D}(z_1,\mu^2)\! \int\limits_0^1\!\! \frac{dz_2}{z_2}D_{j\to \overline{D}}(z_2,\mu^2)\times\nonumber\\
\times\frac{d\sigma(p+p\to i(p_i)+j(p_j)+X)}{dp_{iT} dy_i dp_{jT} dy_j},\label{eq:frag}
\end{eqnarray}
where $D_{i\to D}(z,\mu^2)$ is the fragmentation function (FF) for producing the $D$-meson
from the parton $i$, created at the hard scale $\mu$, the fragmentation parameter $z$ is defined through the relation $p_i=p_D/z$,
with $p_D$ and $p_i$ to be $D$-meson and
$i$-parton four-momenta, correspondingly, and their rapidities $y_D=y_i$. In our calculations we use the LO FFs from Ref.~\cite{FFs}, where the fits of
nonperturbative $D^{0}$, $D^+$, $D^{*+}$, and $D^{+}_{s}$ FFs, both at LO and
NLO in the $\overline{MS}$ factorization scheme, to OPAL data from
LEP1~\cite{OPAL} were performed. These FFs satisfy two desirable
properties: at first, their $\mu$-scaling violation is ruled by
DGLAP evolution equations; at second, they are universal.
It was shown in Ref.~\cite{FFD}, that the significant part of $D$-mesons is produced through the
gluon and charm quark fragmentation only.

At large $\sqrt S$ the
dominant contributions to cross sections of QCD processes gives
multi-Regge kinematics (MRK), where all
particles have limited (not growing with $\sqrt S$) transverse
momenta and are combined into jets with limited invariant mass of
each jet and strongly separated in rapidities. At leading logarithmic approximation of the Balitsky-Fadin-Kuraev-Lipatov (BFKL)
approach~\cite{BFKL} which  gives the description of QCD
scattering amplitudes in this region, where the logarithms of type $(\alpha_s\log(1/x))^n$
are resummed, only gluons can be produced and each jet is actually a
gluon. At next-to-leading logarithmic approximation (NLA) the terms
of $\alpha_s(\alpha_s\log(1/x))^n$
 are collected and a jet can contain a couple of partons (two
gluons or quark-antiquark pair)\ with close rapidities. Such kinematics is called quasi
multi-Regge kinematics (QMRK). Despite of a great number of
contributing Feynman diagrams it turns out that at the Born level in
the MRK amplitudes acquire a simple factorized form. Moreover,
radiative corrections to these amplitudes do not destroy this form,
and this
phenomenon is called gluon Reggeization~\cite{gRegge}. The full set of the induced and effective
vertices of Reggeon-particle interactions together with Feynman rules was written in Refs.~\cite{KTAntonov} and~\cite{LipVyaz}.
The effective action which includes the fields of Reggeized gluons~\cite{KTLipatov}
and Reggeized quarks~\cite{LipVyaz} was introduced and the non-Abelian gauge invariant theory was developed.

The lowest order in $\alpha_S$ parton subprocesses of the PRA in which gluon or $c$-quark pair is produced are:
a quark-antiquark pair production
$
\mathcal {R} + \mathcal {R} \to c + \bar c\label{eq:RRQQ}
$
and the corresponding gluon pair production in QMRK
via two Reggeized gluons fusion $\mathcal {R} + \mathcal {R} \to g+g$.
We present the amplitudes of these processes
and their matrix elements squared in the work~\cite{NSSjets}.

In the $k_T$-factorization, differential cross section for the $2\to
2$ subprocess has the form:
\begin{eqnarray}
\frac{d\sigma}{dy_1dy_2dp_{1T}dp_{2T}d\Delta\phi}(p + p \to c(p_1)+\bar
c(p_2) + X)= \frac{p_{1T}p_{2T}}{16 \pi^3} \int d\phi_1
\int dt_1\times \nonumber \\
\times \Phi(x_1,t_1,\mu^2) \Phi(x_2,t_2,\mu^2)
\frac{\overline{|{\cal M}(\mathcal{R} + \mathcal{R} \to c + \bar
c)|^2}}{(x_1x_2 S)^2}, \label{eq:QMRKcc}
\end{eqnarray}
where $\phi_1$ is the azimuthal angle between ${\bf p}_T$ and ${\bf
q}_{1T}$ and $\Delta\phi$ is the
azimuthal angle between ${\bf p}_{1T}$ and ${\bf p}_{2T}$. The unintegrated over transverse momenta parton distribution functions
 (UPDFs) $\Phi(x,t,\mu^2)$ depend on Reggeon transverse momentum
${\bf q}_T$ while its virtuality is $t=-|{\bf q}_T|^2$.  The UPDFs
are defined to be related with collinear ones through the equation $xG(x,\mu^2)=\int^{\mu^2}dt \Phi(x,t,\mu^2)$.
We obtain the UPDFs using the
model of Kimber,
  Martin and Ryskin~\cite{KMR,Watt} where the transverse momentum of a parton in the initial state of the hard scattering comes entirely
  from the last step of evolution cascade, and the parton radiated at the last step is ordered in rapidity with the particles produced
  in the hard subprocess.

\section{Results}
\label{sec:three}
\begin{figure}[ht]
\begin{center}
\includegraphics[width=0.5\textwidth, origin=c, clip=]{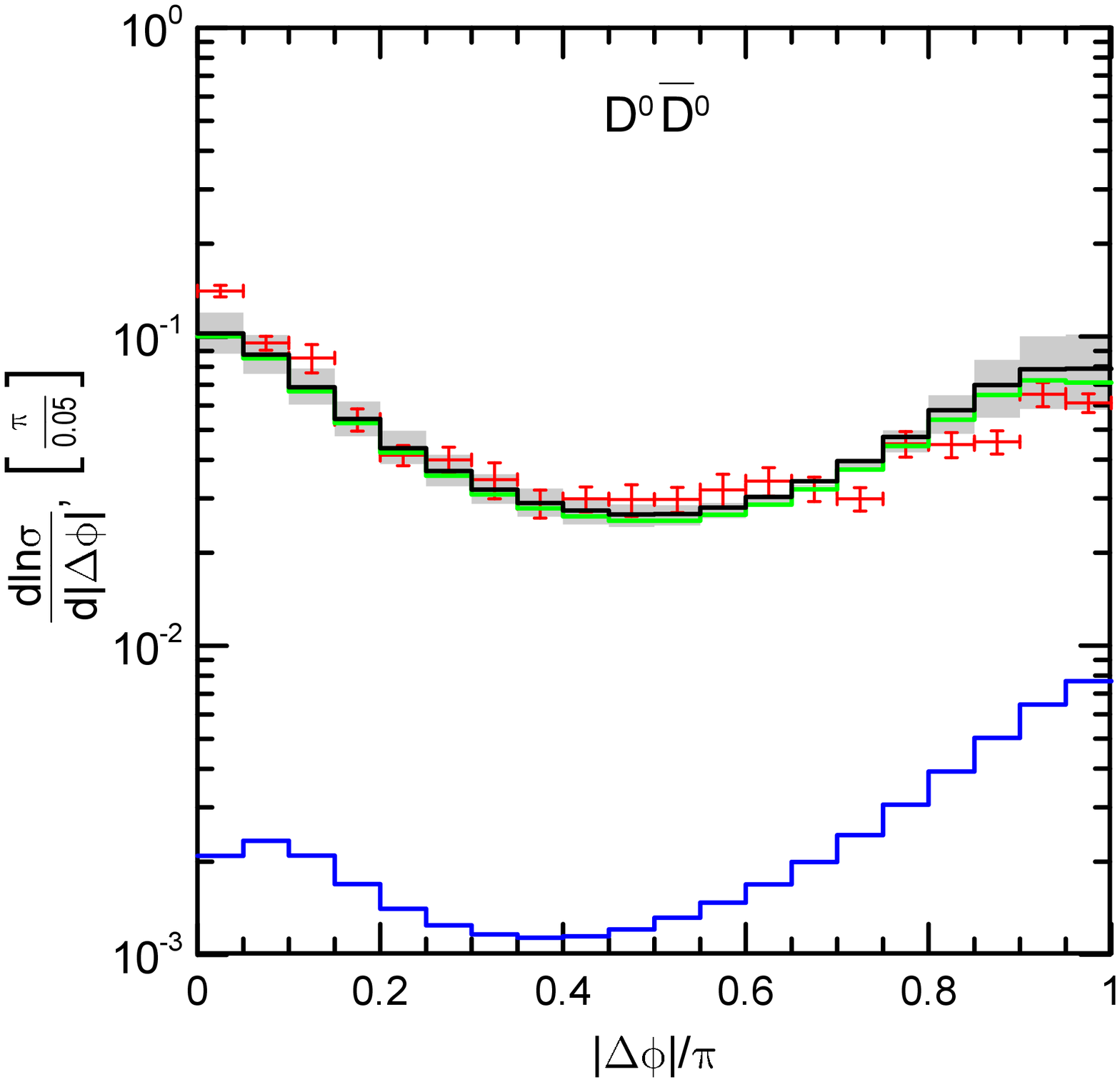}\includegraphics[width=0.5\textwidth, origin=c, clip=]{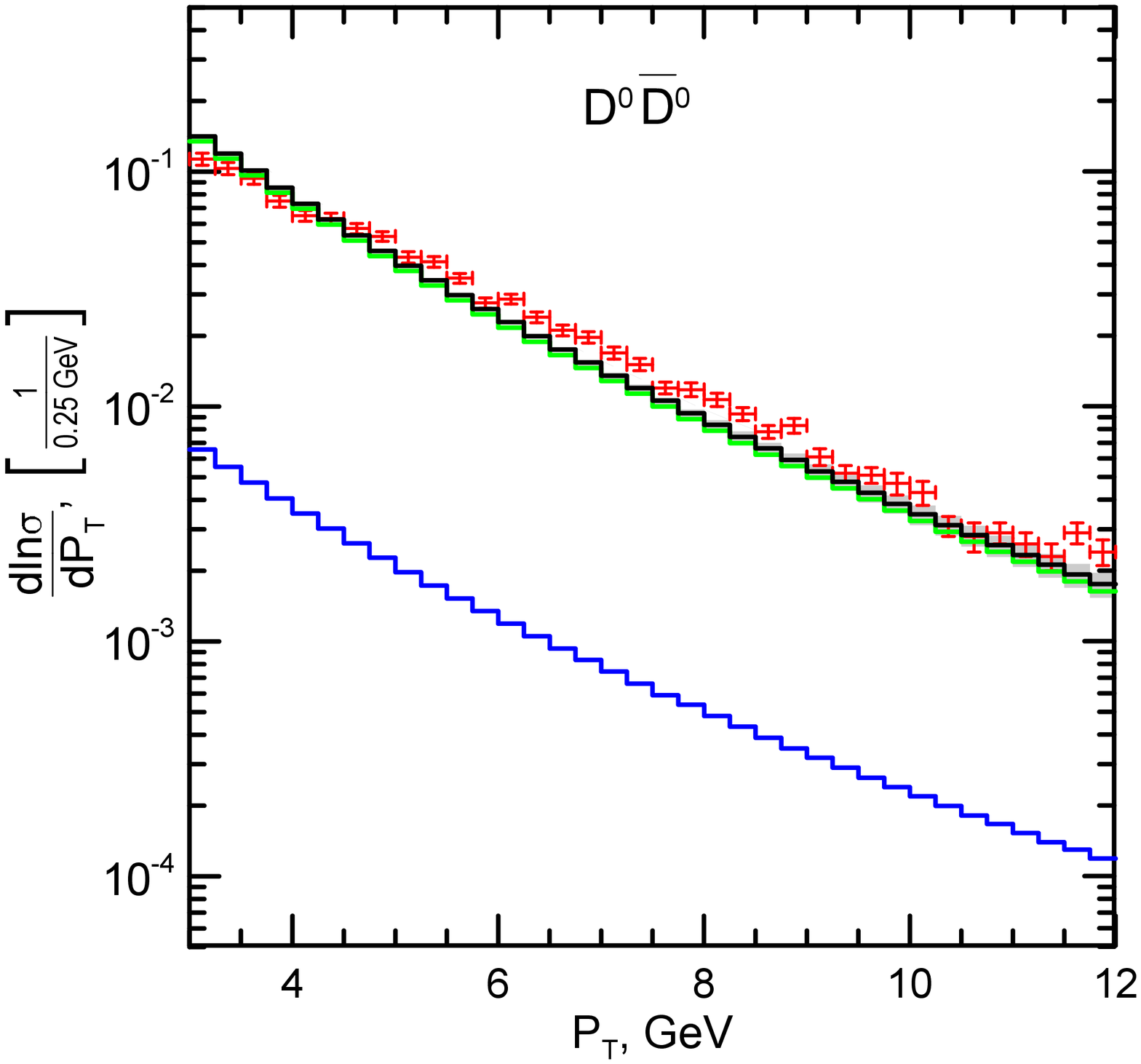}
\includegraphics[width=0.5\textwidth, origin=c, clip=]{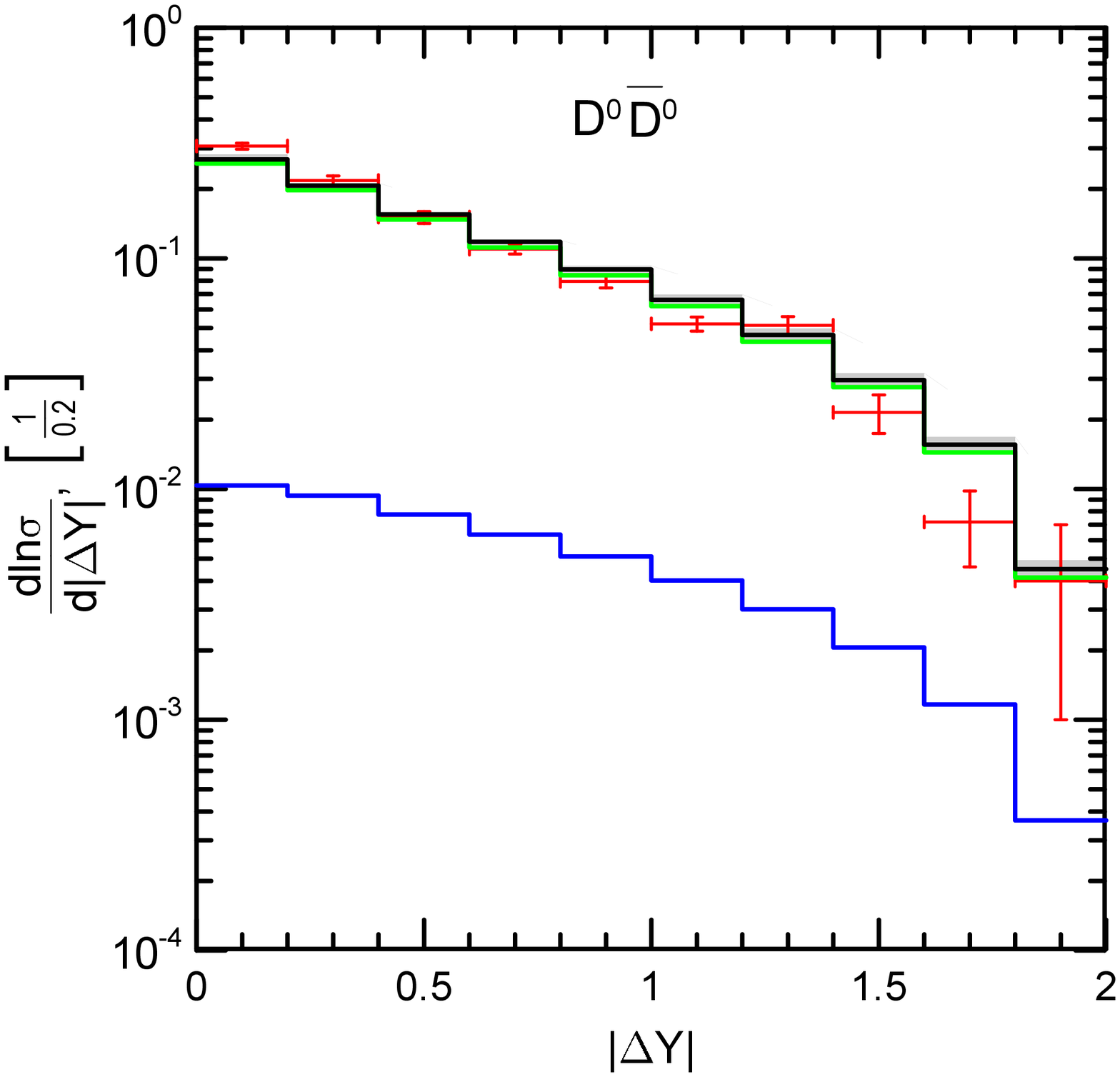}\includegraphics[width=0.5\textwidth, origin=c, clip=]{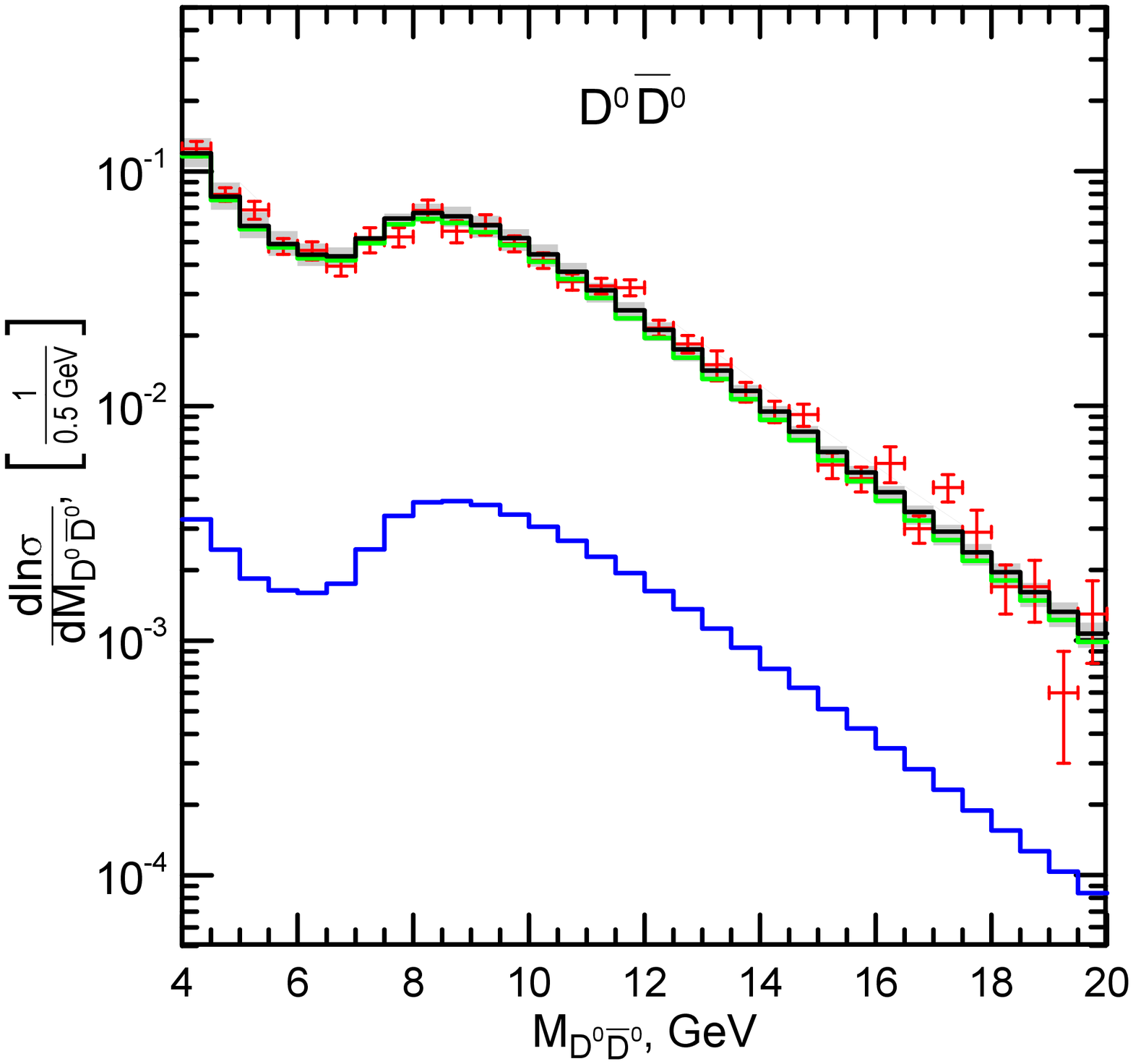}
\end{center}
\caption{Distributions in azimuthal angle between $D^0$ and $\overline{D^0}$ mesons (left-top); in transverse momentum of $D^0$ (right-top); in rapidity distance between $D^0$ and $\overline{D^0}$ (left-bottom) and in $D^0\overline{D^0}$ invariant mass (right-bottom) at LHCb, $\sqrt S=7$~TeV. Green line represents the contribution of $c$-quark fragmentation, blue line -- the gluon-fragmentation contribution, solid line is their sum. The LHCb data are from the Ref.~\cite{LHCb}.
\label{fig:DD1}}
\end{figure}

\begin{figure}[ht]
\begin{center}
\includegraphics[width=0.5\textwidth, origin=c, clip=]{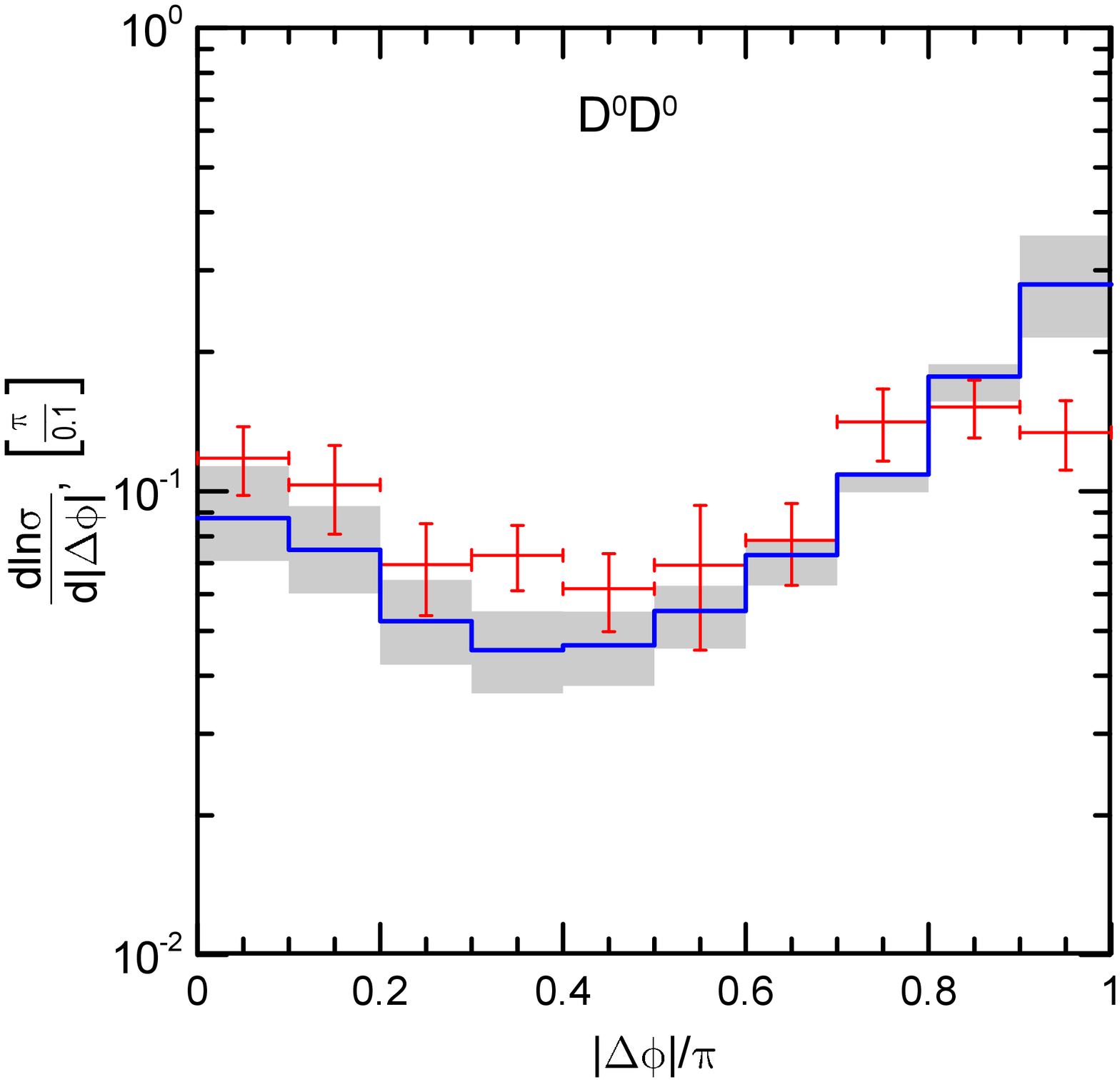}\includegraphics[width=0.5\textwidth, origin=c, clip=]{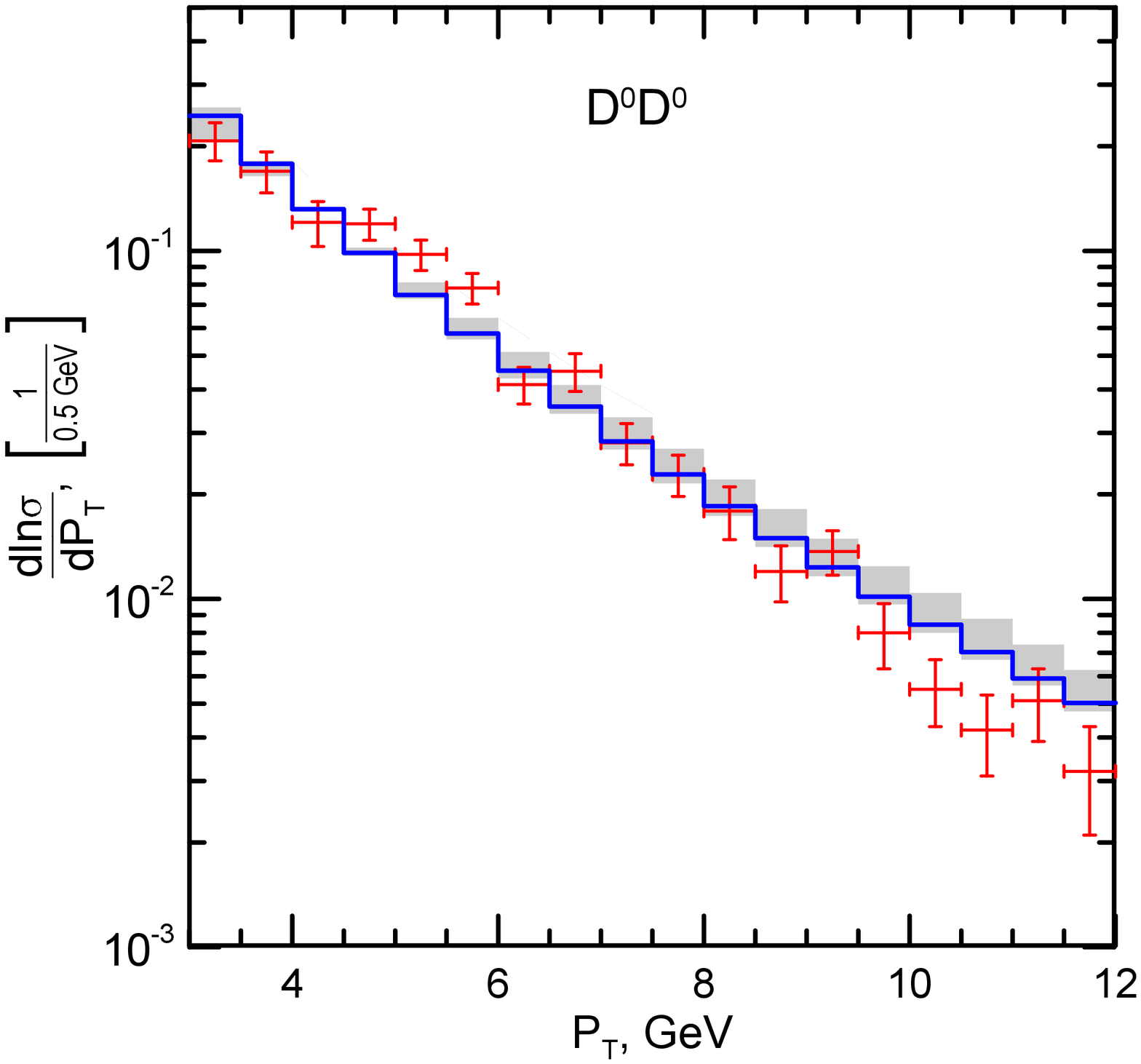}
\includegraphics[width=0.5\textwidth, origin=c, clip=]{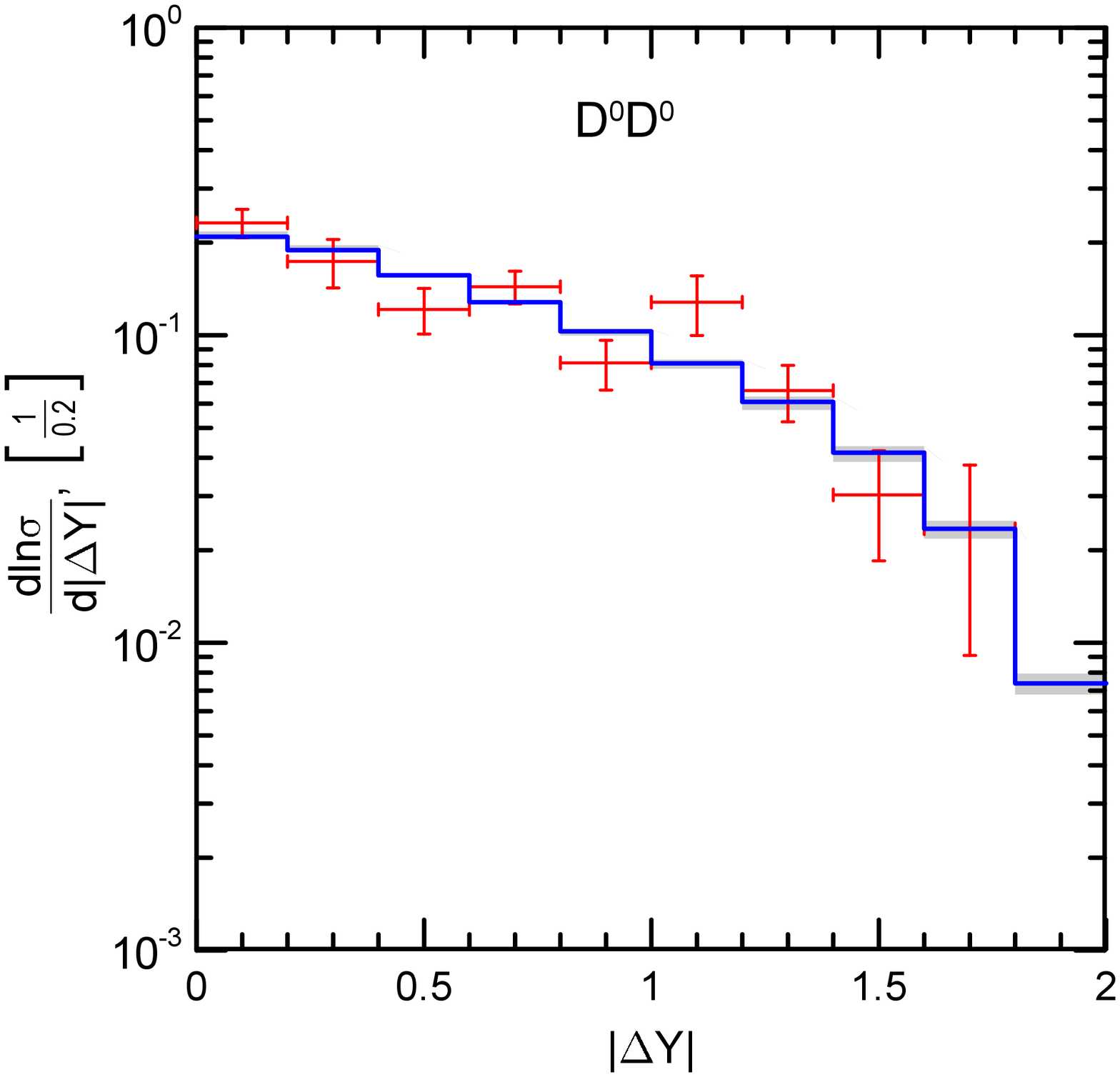}\includegraphics[width=0.5\textwidth, origin=c, clip=]{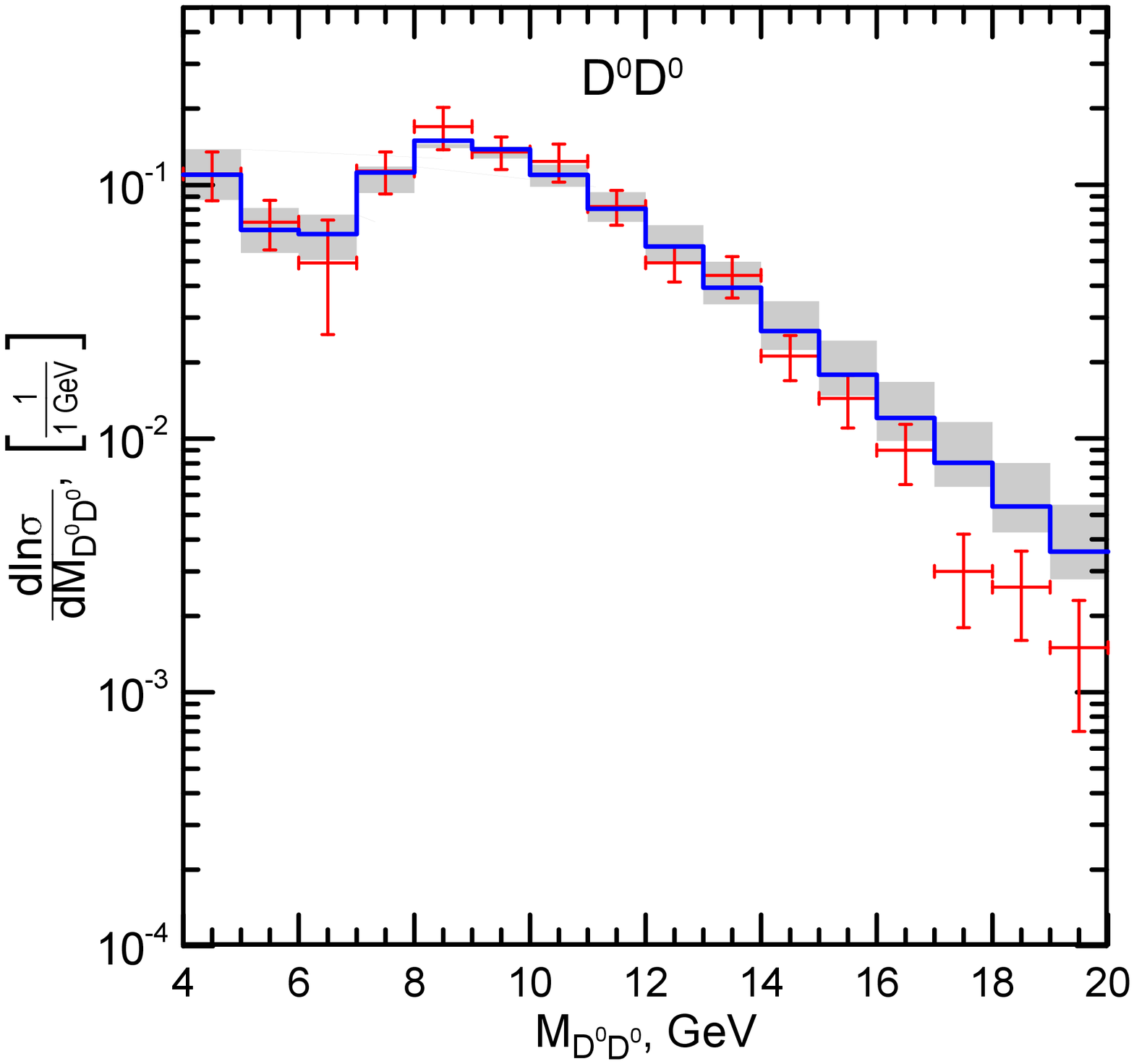}
\end{center}
\caption{Distributions in azimuthal angle between $D^0$ and $D^0$ mesons (left-top); in transverse momentum of $D^0$ (right-top); in rapidity distance between $D^0$ and $D^0$ (left-bottom) and in $D^0D^0$ invariant mass (right-bottom) at LHCb, $\sqrt S=7$~TeV. The blue line represents the gluon-fragmentation contribution. The LHCb data are from the Ref.~\cite{LHCb}.
\label{fig:DD2}}
\end{figure}

We compare the experimental data for $D\overline{D}$-mesons produced at the collision energy of $\sqrt S=7$~TeV at rapidity range
$2.0<y<4.0$~\cite{LHCb}, with our
predictions in the LO of the PRA, in the
Fig.~\ref{fig:DD1}.
The green lines represent contributions of the $c$-quark fragmentation while blue lines correspond to the gluon
ones, and their sum
is shown as a solid line. A theoretical uncertainty
is estimated by varying factorization and
renormalization scales between $1/2 m_T$ and $2 m_T$
around their central value of $m_T$, the transverse mass of a
fragmenting parton. The resulting uncertainty is depicted in the
figures by shaded bands.
The analogous comparison of
the recent data from the LHCb at $\sqrt S=7$~TeV for $DD$ mesons~\cite{LHCb} is presented in the Fig.~\ref{fig:DD2}. Here we take into account only $\mathcal {R} + \mathcal {R} \to g+g$ contribution because in final state we have only $D$ and $D$ mesons, but don't have $\overline{D}$ mesons.
We find a good agreement between our
predictions and experimental data for the different distributions of
$D$-meson pairs within experimental
 and theoretical uncertainties.

\section{Conclusions}
\label{sec:four}

In the present work we performed the study of $D$-meson pair production in proton-proton collisions at the region of large rapidities as it is implemented for the LHCb detector in the framework of Parton Reggeization Approach. Here we take into account all the hard-scattering parton subprocesses appearing at the LO with Reggeized gluons in the initial state. To describe the hard scattering stage we use the fully gauge invariant amplitudes introduced in the works of L.~N.~Lipatov and co-authors. The distributions of initial partons are taken in the form of UPDFs proposed by Kimber, Martin and Ryskin, and the way of their definition is ideologically related to the above-mentioned amplitudes.
 We obtained a good agreement of our results for pair $D$-meson production comparing with experimental data from the LHC,
  especially at large transverse momenta. The achieved degree of agreement is the same as the one obtained
   by NLO calculations in
  the conventional collinear parton model. We also have a good description of the azimuthal distributions at the LO of the PRA in contrast to the collinear parton model. We can see that pair $DD$ production is described very well just by the subprocess $\mathcal {R} + \mathcal {R} \to g+g$. This fact in addition to result of $D\overline{D}$ production indicates that there is no any necessity to involve the hypothesis of double parton scattering~\cite{Szczurek}, see also talk of Prof.~Szczurek~\cite{DIS16}. We describe
  $D$-meson production without any free parameters or auxiliary approximations.

\end{document}